# Quantum State Tomography with a Single Observable

Dikla Oren[1], Maor Mutzafi[1], Yonina C. Eldar[2], Mordechai Segev[1]

[1]Physics Department and Solid State Institute, Technion, 32000 Haifa, Israel

[2] Electrical Engineering Department, Technion, 32000 Haifa, Israel

**Quantum information has been drawing a wealth of research in recent years, shedding light on questions at the heart of quantum mechanics**[1–5]**, as well as advancing fields such as complexity theory**[6–10]**, cryptography**[6]**, key distribution**[11]**, and chemistry**[12]**. These fundamental and applied aspects of quantum information rely on a crucial issue: the ability to characterize a quantum state from measurements, through a process called Quantum State Tomography (QST). However, QST requires a large number of measurements, each derived from a different physical observable corresponding to a different experimental setup. Unfortunately, changing the setup results in unwanted changes to the data, prolongs the measurement and impairs the assumptions that are always made about the stationarity of the noise.** *Here, we propose to overcome these drawbacks by performing QST with a single observable*. **A single observable can often be realized by a single setup, thus considerably reducing the experimental effort. In general, measurements of a single observable do not hold enough information to recover the quantum state. We overcome this lack of information by relying on concepts inspired by Compressed Sensing (CS)**[13,14]**, exploiting the fact that the sought state – in many applications of quantum information - is close to a pure state (and thus has low rank). Additionally, we increase the system dimension by adding an ancilla that couples to information evolving in the system, thereby providing more measurements, enabling the recovery of the original quantum state from a single-observable measurements.**

**We demonstrate our approach on multi-photon states by recovering structured quantum states from a single observable, in a single experimental setup. We further show how this approach can be used to recover quantum states without number-resolving detectors.**

The fields of quantum information and quantum computation have been attracting considerable interest in recent years. From fundamental quantum mechanics (the measurement process[1], description[2] and role[3] of entanglement, and information content of quantum measurements[4]) to applied physics (molecular energies calculations[12], experimental quantum key distribution[15]), quantum information and computation have helped raise questions and aid in the investigations of many topics in physics and related fields. Rooted in Feynman's ideas[16], the concept of a quantum computer has scaled and evolved, yielding programmable computation units[17,18]. In particular, optical quantum computation, based on the concepts of KLM[19] and cluster states[20–22], has recently witnessed experimental realizations of larger and more complex systems in terms of photon number[22,23]. Alternatives to the common quantum circuit model have also been suggested[20,24], among them a recent linear non-universal scheme for quantum computing[10,25–28], which is thought to yield a real quantum advantage over classical information processing already with the near-future technology.

Whether exploring fundamental quantum mechanics, or advancing quantum computation, the ability to characterize the quantum state from measurements is principal to the fields of quantum information and computation. As the fundamental description of a quantum system is given by the density matrix, the characterization amounts to identifying all elements of this matrix. The density matrix, a positive semi-definite, trace-normalized matrix, allows prediction of every experimental result, thus providing the full description of a quantum system. To recover the density matrix, Quantum State Tomography (QST) is usually performed. In this process, the density matrix is measured by a large number of observables, each corresponding to a different experimental setup. Each observable is associated with a Hermitian operator, whose eigenvalues are the outcome of the measurements, whereas their probabilities are derived from the eigenstates. By choosing a suitable set of observables, known as a "tomographically complete set of observables", the elements of the density matrix can be recovered from the measurements. The process of QST requires a large number of measurements, realized in multiple experimental setups. Generally, for a system of dimension $d$, the density matrix is described by $d^2 - 1$ real parameters. Since each observable yields at most $d$ measurement outcomes, a tomographically complete set of observables consists of $d$ observables, each corresponding to a different experimental setup. Naturally, the need to change the physical setup (even by rotating a waveplate or a polarizer) increases the duration of the experiment, and hampers the integrity of the state. Perhaps even more importantly, the need to carry out many variations of the experimental setup often hurts the assumptions made on the noise. That is, in all quantum optics experiments the flux of entangled states is low, hence the integration times are long (minutes and longer). Consequently, the detection process always assumes that the noise does not change during the experiment. Clearly, varying the experimental setup due to the need to measure multiple observables, as well as having to carry out a large number of measurements, hampers these assumptions. Altogether, it would be highly desirable to be able to recover quantum states using a single observable, in a single experimental setup. However, as we explain below, measurements obtained from a single observable are known to be insufficient for full QST. Consequently, the recovery of quantum states from a single observable has never been realized experimentally.

Here, *we propose recovering the quantum state using a single observable, corresponding to a single experimental setup*. Since a single observable of dimension $d$ can only yield $d$ measurement outcomes, whereas $d^2 - 1$ are needed for complete recovery of the density matrix, relying on a single observable implies that information will be missing. To overcome this lack of information while keeping the measurement number to the necessary minimum, we may rely on prior information, which should be as generic as possible so as to maintain the applicability of our approach to a large class of settings. With this in mind, we exploit generic prior information: that the sought quantum states are close to pure states. This prior is natural in quantum information, because many of its applications deal with mapping pure states onto other pure states. Of course, in a physical system the states are not ideally pure – due to noise in the generation, manipulation and detection of the quantum states. Nonetheless, the states are in most cases still close to pure states, hence the density matrix has a small number of nonzero eigenvalues, i.e., the eigenvalues are sparse.

Exploiting sparsity is at the heart of the field of Compressed Sensing (CS)[13,14,29], a very active area of research within signal processing, which enables reconstruction of information from incomplete measurements by exploiting sparse priors. More recently, CS has been brought into the quantum domain for the purpose of reducing the number of measurements necessary in QST[30] and in quantum process tomography[31], enabling much more efficient tomography. It was further used in wavefunction measurements[32,33], measurements of complementary observables[32], weak measurements[34,35], characterization of incoherent light[36], holography[37] and ghost imaging[38]. The general concept of using sparsity to solve underdetermined inverse problems has opened the door for a wide range of applications in various fields, ranging from sub-Nyquist sampling[39], sub-wavelength imaging[40–43], phase retrieval[41,42,44–46], to Ankylography[47], Ptychography[48] and quantum state recovery from low-order correlations[49].

The sparsity naturally arising in quantum information comes in the form of the quantum states, which, for most applications, tend to be pure states or close to pure states. Such states are of interest from a variety of reasons. First, pure states have zero entropy. Thus, as a random variable, they contain the most information and many theoretical results in quantum information apply to them. Second, the purity of a state is invariant under unitary transformations. The purity is defined as $\mathcal{P}(\rho)=\text{Tr}(\rho^2)$ ($\rho$ being the density matrix), with $\mathcal{P}=1$ for a pure state, and $\mathcal{P}<1$ for mixed states. Thus, under time-evolution of a closed system pure states always remain pure states. These are the reasons that most applications of quantum information ideally deal with mapping pure states onto other pure states. Of course, in a realistic experimental scenario, uncertainties and imperfections do exist, and the system is not always closed. Thus, the resulting states in an experimental system are not perfectly pure, but they can often be described by states which are close to pure states. Finally, some quantum channels, describing noise processes and open system evolution, map pure states to states which can be approximated by relatively pure states. In the language of the density matrix, a pure state is described by a Rank-1 density matrix, whereas a relatively pure state is described by a low-rank density matrix, having a small number of nonzero (sparse) eigenvalues, relative to the system dimension. Accordingly, a state which can be approximated by a relatively pure state has a small number of significant eigenvalues. All these states fall under the category of sparse (or compressible) states, and are addressed by our method.

An essential requirement for CS recovery to work well is that each measurement carries information. This is achieved by performing measurements in a basis which is least correlated (so-called the "least coherent" in the language of CS) with the basis providing the sparse representation. In the field of optics, for example, two such bases naturally occur in the form of real space and Fourier space. For a sparse signal in real space, measurements performed in Fourier space are good for CS, and vice versa. Thus, in the spirit of CS, we introduce mixing between the channels in the system. In the context of indistinguishable photons in discrete spatial modes, the mixing is realized by a random, linear coupler, which can be experimentally realized by beamsplitters[50] or by integrated photonics[17,27,51].

However, measurements taken with a single observable do not contain enough information for recovering the state of a quantum system (the density matrix). This is because a single observable of dimension $d$ yields at most $d$ different measurement outcomes, whereas even a pure state has more degrees of freedom $(2d - 1)$, let alone relatively pure states, which require even more measurements. To overcome the lack of measurements, we add an ancilla in a known state to the state we wish to recover. This requires a short explanation about ancillas in quantum information.

Using known inputs to a system to improve its probing is a widely used concept in optical detection, from spectral interferometry[52], to optical homodyne detection[53,54] and integrated photonics schemes[55–57]. In the quantum context, consider a quantum system of $m$ channels carrying information, with a coupler that couples (mixes) the information in the channels. An ancilla is the addition of $m'$ new channels with known inputs (say, zero input) but the evolution in the system couples these channels with the channels of the original system. The total number of channels where the measurements take place is therefore $M = m + m'$. The input state of the extended system is a tensor product of the original state $\rho_0$ and the ancilla state $\rho_{\text{ancilla}}$, such that $\rho_{\text{in}} = \rho_0 \otimes \rho_{\text{ancilla}}$. Because the input is a simple product state, its mutual information ($\mathcal{I}(A:B) = \mathcal{S}(A) + \mathcal{S}(B) - \mathcal{S}(A,B)$, where $\mathcal{S}(\rho) = -\text{Tr}(\rho \log \rho)$ is the von Neumann entropy) is zero:

$\mathcal{S}(\rho_0 \otimes \rho_{\text{ancilla}}) = \mathcal{S}(\rho_0) + \mathcal{S}(\rho_{\text{ancilla}}) \Rightarrow \mathcal{I}_{in}(\text{original:ancilla}) = 0$. However, after the evolution, the state is $\rho_{\text{out}} = \mathcal{U}\rho_0 \otimes \rho_{\text{ancilla}}\mathcal{U}^\dagger$, where $\mathcal{U}$ is the evolution operator in the system. Whenever the evolution couples between the ancilla and the original system, it is not a tensor product anymore, that is, $\mathcal{U} \neq \mathcal{U}_{\text{original}} \otimes \mathcal{U}_{\text{ancilla}}$, hence $\rho_{\text{out}} \neq \rho_0^{\text{out}} \otimes \rho_{\text{ancilla}}^{\text{out}}$, hence the mutual information at the output is always larger than zero. Physically, this means that the ancilla effectively increases the number of measurements (see Supplementary Information). In our system specifically, the dimension of the ancilla should be large enough so as to account for the missing information to facilitate the recovery of the input state from a single observable. In Photonics, the ancilla is conveniently realized by adding vacuum ports at the input. Surely, increasing the system dimension to the full $d^2 - 1$ degrees of freedom (for a generic density matrix of dimension $d$) is possible, such that a single observable accounts for all the measurements required for QST. However, for large values of $d$, the number of required measurements and the number of required ancilla channels for full QST can be very large. Here, using the prior knowledge that the input state is sparse allows us to considerably reduce the required dimensional increase of the ancilla.

With the notions of sparsity, the addition of the ancilla and the mixing between the degrees of freedom in mind, we can formulate the problem. Our goal is to recover a density matrix $\rho_0$ of dimension $d$ and rank $r$ (unknown but small relative to $d$) from measurements of a single observable $A$. Consider a system of $N$ photons in $m$ ports (Fig. 1(a), with $N = 3, m = 4$), having dimension $d = \binom{m-1+N}{N}$ (in Fig. 1(a), $d = 20$). We add vacuum ports (Fig. 1(b)), realizing the ancilla in the state $|\psi\rangle_{\text{ancilla}} = |0\rangle$, such that the total number of ports is $M > m$ ($M - m$ vacuum ports are added, in Fig. 1(b,c) $M = 8, M - m = 4$). The dimension of the system is now $D = \binom{M-1+N}{N} > d$, and the state of the joint system is described by the $D$-dimensional density matrix $\rho = \rho_0 \otimes \rho_{\text{ancilla}} = \rho_0 \otimes |0\rangle\langle 0|$. The mixing is realized by a linear, random coupler $U$ of $M$ ports,

introduced between the input and the number resolving detectors at the output (Fig. 1(c)), causing the state to evolve according to

$$\rho_0 \otimes |0\rangle\langle 0| \mapsto \mathcal{U}(\rho_0 \otimes |0\rangle\langle 0|)\mathcal{U}^\dagger, \qquad (1)$$

where $\mathcal{U} \in \mathrm{U}(D)$ is the evolution of the entire system, original state and ancilla, dictated by the coupler $U \in \mathrm{U}(M)$ such that $\mathcal{U} \neq \mathcal{U}_{\mathrm{original}} \otimes \mathcal{U}_{\mathrm{ancilla}}$. See the Supplementary Information for further details. The measurements performed are correlation measurements of $N$ photons. This set of measurements, along with the linear coupler, define the observable we use:

$$A = \sum_i i \mathcal{U}^\dagger |\{n\}^i\rangle\langle\{n\}^i| \mathcal{U} \in \mathbb{C}^{D \times D}, \qquad (2)$$

Where $A^\dagger = A$ and $|\{n\}^i\rangle = |n_1^i n_2^i \cdots n_M^i\rangle$ is the $i$th Fock state with $n_q^i$ photons in port $q$ (see details in the Supplementary Information). Experimentally, these measurements describe the $N$-fold correlation measurements after the linear coupler. The problem now translates into finding $\rho_0 \in \mathbb{C}^{d \times d}$, a positive semidefinite matrix $\rho_0^\dagger = \rho_0, \rho_0 \geq 0$ (an Hermitian matrix with non-negative eigenvalues) with unit trace $\mathrm{Tr}(\rho_0) = 1$, having the lowest rank and conforming to the measurements

$$y_i = \mathrm{Tr}(\rho \mathcal{U}^\dagger |\{n\}^i\rangle\langle\{n\}^i| \mathcal{U}) = \langle\{n\}^i| \mathcal{U}\rho\mathcal{U}^\dagger |\{n\}^i\rangle, i \in \{1, \dots, D\}. \qquad (3)$$

We emphasize that the number of measurements here is $D = \binom{M-1+N}{N}$. However, since they are all derived from a single realization of the coupler, all of these measurements can be realized in a single experimental setup.

The density matrix we wish to find is the solution to the problem

$$\min_{\rho_0} \mathrm{rank}(\rho_0)$$
$$\text{subject to } \rho_0^\dagger = \rho_0, \rho_0 \geq 0, \mathrm{Tr}(\rho_0) = 1$$

$$|\text{Tr}(\rho_0 \otimes |0\rangle\langle 0|A_i) - y_i| \leq \epsilon, i = 1, \dots, D. \quad (4)$$

Here, $A_i = \mathcal{U}^\dagger|\{n\}^i\rangle\langle\{n\}^i|\mathcal{U}$ is the spectral decomposition of the single observable (eq. (2)). This optimization problem is related to the Matrix Completion problem[58–60] with the additional constraints stemming from the physical nature of the object we wish to recover and the measurements derived from the single observable. This problem is not convex, since the rank objective is not convex. Thus, to find the density matrix $\rho_0$, we utilize the LogDet[61] approach, in which the non-convex rank is replaced by the logarithm of the matrix determinant, which in turn is linearized to yield the following iterative algorithm

$$\min_{X_k} \text{Tr}(X_{k-1} + \delta I)^{-1} X_k$$
$$\text{subject to } X_k \geq 0, \text{Tr}(X_k) = 1, X_k^\dagger = X_k,$$
$$|\text{Tr}(X_k \otimes |0\rangle\langle 0|A_i) - y_i| \leq \epsilon, \ i = 1, \dots, D. \quad (5)$$

Here, in iteration $k$, we look for the matrix $X_k$, where $X_{k-1}$ is the solution of the previous iteration and $\delta$ is a small regularization parameter. The parameter $\epsilon$ is related to the measurement noise. Other methods can be utilized as well, see details in the Supplementary Information.

The information evolves in our setting is as follows. Consider an ensemble of quantum states of varying ranks $r \in \{1, \dots, d\}$. Each state undergoes a depolarization channel $\rho_0 \mapsto (1-\mu)\rho_0 + \frac{\mu}{d}I_d$, realizing noise in the state. The resulting state has $r$ significant eigenvalues. Then, the density matrix evolves in the linear coupler, which has a large dimension and provides mixing between the ports. The information propagates to the output ports, which is where the measurements of the single observable are taken. The measured single observable includes noise added to it. This is the "hardware" defining our system.

Next, we demonstrate the power of our scheme to recover quantum states of rank $r$ (unknown but small relative to the dimension of the system). The input to our recovery procedure is the noisy single-observable measurements $y_i, i \in \{1, ..., D\}$. The algorithm (described in the Supplementary Information) results in the recovered density matrix $\rho_{rec}$. To evaluate the performance of the recovery, we compare $\rho_{rec}$ to the original density matrix $\rho_0$, by means of the fidelity between the two states $\mathcal{F}(\rho_0, \rho_{rec}) = \text{Tr}\sqrt{\rho_0^{1/2} \rho_{rec} \rho_0^{1/2}}$.

An example of the original and recovered density matrices is presented in Fig. 2(a). The density matrix of rank 2 describes 3 photons in 3 input ports. The ancilla used consists of 4 vacuum ports, yielding a total of 7 output ports. The original and recovered density matrices match very well, with a fidelity of 0.96. To test the performance of our methodology, we generate an ensemble of mixed density matrices of various ranks. The matrices are sampled from a product measure of the eigenvalues and eigenvectors, where the uniform measures on the unit simplex and the Unitary group are used, respectively (see details in the Supplementary Information). The recovery fidelity without measurement noise, averaged over 200 random realizations of the density matrix for each rank and 10 realizations of the random coupler, is shown in Fig. 2(b), solid lines. For each recovery, measurements from a single coupler are used, as the different realizations are employed for the sake of averaging (see Supplementary Information). The different solid curves correspond to a varying number of output ports $M \in \{7,9,11\}$. As expected, the fidelity is very high for low ranks, describing states which are relatively close to pure states. The fidelity grows with the increase of the system dimension through the ancilla (which also increases the dimension of the single observable). The fraction of measurements used, out of the total number of measurements required for full QST, is 21,42,71% for $M = 7,9,11$, respectively. For example, for Rank 2

(almost pure states), we can recover the quantum state with only 21% of the total measurements, while for Rank 6 we would need 71%., in this small system.

The dotted lines in Fig. 2(b) describe the average recovery fidelity in a noisy scenario. Here, depolarization noise of 2% is added to the state, and measurement noise of $25 dB$ is added to the measurements. Once again, the recovery from a single setup works well for low rank density matrices describing states which are close to pure states. Importantly, the recovery does not depend on the exact realization of the coupler, as long as it is sampled from the correct distribution (see Supplementary Information for details). Motivated by BosonSampling[10], the theory and experimental realization of such Haar random linear optical couplers have been developed significantly[51,62]. In Fig. 2(c), the mean recovery fidelity is shown for a state of a larger dimension. Here, the density matrices describe $N = 3$ photons in $m = 7$ input ports, $M = 16$ output ports. The dimension of the system is $d = 84$, while the number of measurements is $D = 816$, meaning that high fidelity reconstruction is achieved with only 11% of the measurements required for full QST. In Fig. 2(d), we compare the recovery fidelity of a Haar random coupler to that of a simpler coupler, consisting of uniform, evanescently coupled waveguides with nearest-neighbor coupling only. The simple waveguide array (1D photonic lattice) offers some mixing between the degrees of freedom, but at a level considerably lower than the Haar random coupler. As evident in the figure, the simpler coupler fails to allow the recovery from a single observable, in a single setup, whereas the Haar coupler performs perfectly up to Rank 6.

As seen in the results for a large dimension (Fig. 2(c)) comparing to the results in Fig. 2(b), as the dimension of the original state increases, low-rank states become amenable for recovery from a smaller portion of the measurements required for complete QST (~0.11 of the measurements for $d = 84$ comparing to ~$0.21 - 0.71$ of the measurements for $d = 20$). This result is expected from theory of CS, where for ideal measurement matrices, the number of measurements required to recover a rank-$r$ state is $N_{\text{measurements}} \geq O(rd\,\text{poly}\log d)$[30,63]. In our case, $N_{\text{measurements}} = D = \binom{M-1+N}{N}$, where $M$ is the number of output ports, $N$ is the number of photons. Therefore, we expect that as the dimension increases, the sparsity will enable recovery of the state from a single observable measurements with a smaller portion of the measurements required for full QST.

Following the success of our methodology to recover the quantum states from a single observable, we wish to tackle another problem having to do with recovering quantum states from partial measurements: recovering the quantum state using click detectors (i.e., detectors that cannot resolve the number of photons detected). Generally, photon number resolving detectors are required for preforming QST of multi-photon states. However, such detectors often exhibit poor performance (primarily in terms of sensitivity and fall times, which affects the sampling rate), and currently the majority of quantum optics experiments are carried out with click detectors. One possible avenue for optical QST with click photodetectors is to mimic their functionality by the addition of beamsplitters[64]. However, this method introduces further losses to the system and is not scalable to larger number of channels and photons. A different approach has recently been shown to obtain the counting statistics of light[55,56].

As we now show, our scheme works even in the extreme case of combining a single-observable with click detectors. That is, we recover the full density matrix of a multi-photon state with a fixed number of photons, from a single observable and without number resolving detectors. For an $N$ photon state in $m$ ports and $m > N$, we use regular click-detectors, detecting the presence of more-than-zero photons in each detection event. We use only the detection events involving $N$ different clicks. In these events, counting and detecting are the same. Now, we use a partial set of measurements: $y_i = \langle\{n\}^i|\mathcal{U}\rho\mathcal{U}^\dagger|\{n\}^i\rangle$, as before, only with $i \in I$ such that $I = \{i \in \{1,...,D\} \mid n_q^i \in \{0,1\} \,\forall q \in \{1,...,M\}\}$ and $|I| = \binom{M}{N}$. Naturally, these events do not contain all the information[65] needed for quantum state reconstruction. However, by using our scheme, we can overcome this deficit and recover the full density matrix from such very partial measurements.

Figure 3(a) shows an example of a rank-2 state describing $N = 3$ photons in $m = 3$ input ports. Noise of $25dB$ is added to the measurements, and the state is recovered from a single observable, without number resolving detectors. In Fig. 3(b), the recovery fidelity versus the rank of the density matrix describing $N = 3$ photons in $m = 4$ input ports, $M = 11$ output ports and measurement noise of $25dB$, without number resolving detectors. Low rank states are recovered from the noisy measurements, in a single setup, without number resolving detectors.

In the state recovery using click detectors, the number of measurements depends on the system characteristics as $N_{\text{measurements}} = \binom{M}{N}$, where $M$ is the number of output ports, $N$ the number of photons. Here as well, we expect that the sparsity will reduce the portion of measurements required to recover the state with click detectors as the dimension of the original state $d$ increases.

Finally, the two aspects of our approach, namely preforming tomography with a single observable and in a single setup, are augmented by a third point of view. Often, QST is formulated in the language of Positive Operator Valued Measurements (POVM)[66], which are generalized quantum measurements. In the POVM language, measurements are described not by observables, but rather by a set of positive semidefinite operators $\{E_i\}$, $E_i^\dagger = E_i, E_i \geq 0$ which sum up to the identity $\sum_i E_i = I$. The probability for each result is given by $p(i|\rho) = \text{Tr}(\rho E_i)$. This is a generalization of the observables-related measurements, since the POVM elements need not be orthogonal. Our method can be thought of as an efficient approach to Neumark theorem[67,68], where an ancilla is used to realize general quantum measurements (POVM) using projective measurements. See the Supplementary Information for further details.

In conclusion, we showed that the generic prior knowledge of having a low rank density matrix (sparsity) can be used to recover the complete quantum state from measurements of a single observable, often corresponding to a single experimental setup, in an efficient manner. This is achieved by adding an ancilla to the original state and an introduction of a random linear coupler between the input state and the measurements. We further used these ideas to recover the complete density matrix of a state (with a fixed number of photons), with click detectors. We have shown how the main ideas of the scheme can be implemented in a system of $N$ photons in $m$ input modes, however these ideas can be extended to any system supporting the addition of an ancilla and enough mixing between the degrees of freedom in the form of interactions. A natural development of our scheme would be to try to optimize the linear coupler such that it is ideal in the sense of CS. A further development is addition of mixing between fixed photon number subspaces in the Hilbert space, aiding in the recovery of density matrices describing multi-photon states with a varying photon number, without number resolving detectors.

# References

1. Peres, A. & Wootters, W. K. Optimal detection of quantum information. *Phys. Rev. Lett.* **66,** 1119–1122 (1991).

2. Horodecki, R., Horodecki, P., Horodecki, M. & Horodecki, K. Quantum entanglement. *Rev. Mod. Phys.* **81,** 865–942 (2009).

3. Vidal, G. Efficient Classical Simulation of Slightly Entangled Quantum Computations. *Phys. Rev. Lett.* **91,** 147902 (2003).

4. Holevo, A. S. Bounds for the quantity of information transmitted by a quantum communication channel. *Probl. Peredachi Informatsii* **9,** 3–11 (1973).

5. Bennett, C. H. *et al.* Teleporting an unknown quantum state via dual classical and Einstein-Podolsky-Rosen channels. *Phys. Rev. Lett.* **70,** 1895–1899 (1993).

6. Shor, P. W. Algorithms for quantum computation: discrete logarithms and factoring. in *Proceedings 35th Annual Symposium on Foundations of Computer Science* 124–134 (IEEE Comput. Soc. Press). doi:10.1109/SFCS.1994.365700

7. Simon, D. R. On the Power of Quantum Computation. *SIAM J. Comput.* **26,** 1474–1483 (1997).

8. Bennett, C. H., Bernstein, E., Brassard, G. & Vazirani, U. Strengths and Weaknesses of Quantum Computing. *SIAM J. Comput.* **26,** 1510–1523 (1997).

9. Knill, E. & Laflamme, R. Power of One Bit of Quantum Information. *Phys. Rev. Lett.* **81,** 5672–5675 (1998).

10. Aaronson, S. & Arkhipov, A. The computational complexity of linear optics. in *Proceedings of the 43rd annual ACM symposium on Theory of computing - STOC '11* 333 (ACM Press, 2011). doi:10.1145/1993636.1993682

11. Bennett, C. H. Quantum cryptography: Public key distribution and coin tossing. in *International Conference on Computer System and Signal Processing, IEEE, 1984* 175–179 (1984).

12. O'Malley, P. J. J. *et al.* Scalable Quantum Simulation of Molecular Energies. *Phys. Rev. X* **6,** 31007 (2016).

13. Candès, E. J. Compressive sampling. in *Proceedings of the international congress of mathematicians* **3,** 1433–1452 (Madrid, Spain, 2006).

14. Eldar, Y. C. & Kutyniok, G. *Compressed sensing: theory and applications*. (Cambridge University Press, 2012).

15. Zhao, Y., Qi, B., Ma, X., Lo, H.-K. & Qian, L. Experimental Quantum Key Distribution with Decoy States. *Phys. Rev. Lett.* **96,** 70502 (2006).

16. Feynman, R. P. Simulating physics with computers. *Int. J. Theor. Phys.* **21,** 467–488 (1982).

17. Carolan, J. *et al.* Universal linear optics. *Science (80-. ).* **349,** (2015).

18. Debnath, S. *et al.* Demonstration of a small programmable quantum computer with atomic qubits. *Nature* **536,** 63–66 (2016).

19. Knill, E., Laflamme, R. & Milburn, G. J. A scheme for efficient quantum computation with linear optics. *Nature* **409,** 46–52 (2001).

20. Raussendorf, R., Browne, D. E. & Briegel, H. J. Measurement-based quantum computation on cluster states. *Phys. Rev. A* **68,** 22312 (2003).

21. Nielsen, M. A. Optical Quantum Computation Using Cluster States. *Phys. Rev. Lett.* **93,** 40503 (2004).

22. Schwartz, I. *et al.* Deterministic generation of a cluster state of entangled photons. *Science (80-. ).* (2016).

23. Lu, C.-Y. *et al.* Experimental entanglement of six photons in graph states. *Nat. Phys.* **3,** 91–95 (2007).

24. Farhi, E., Goldstone, J., Gutmann, S. & Sipser, M. Quantum Computation by Adiabatic Evolution. (2000).

25. Spring, J. B. *et al.* Boson Sampling on a Photonic Chip. *Science (80-. ).* **339,** (2013).

26. Broome, M. A. *et al.* Photonic Boson Sampling in a Tunable Circuit. *Science (80-. ).* **339,** (2013).

27. Tillmann, M. *et al.* Experimental boson sampling. *Nat Phot.* **7,** 540–544 (2013).

28. Crespi, A. *et al.* Integrated multimode interferometers with arbitrary designs for photonic boson sampling. *Nat Phot.* **7,** 545–549 (2013).

29. Eldar, Y. C. *Sampling theory : beyond bandlimited systems*.

30. Gross, D., Liu, Y.-K., Flammia, S. T., Becker, S. & Eisert, J. Quantum State Tomography via Compressed Sensing. *Phys. Rev. Lett.* **105,** 150401 (2010).

31. Shabani, A. *et al.* Efficient Measurement of Quantum Dynamics via Compressive Sensing. *Phys. Rev. Lett.* **106,** 100401 (2011).

32. Howland, G. A., Schneeloch, J., Lum, D. J. & Howell, J. C. Simultaneous Measurement of Complementary Observables with Compressive Sensing. *Phys. Rev. Lett.* **112,** 253602 (2014).

33. Mirhosseini, M., Magaña-Loaiza, O. S., Hashemi Rafsanjani, S. M. & Boyd, R. W. Compressive Direct Measurement of the Quantum Wave Function. *Phys. Rev. Lett.* **113,** 90402 (2014).

34. Howland, G. A., Lum, D. J. & Howell, J. C. Compressive wavefront sensing with weak values. *Opt. Express* **22,** 18870 (2014).

35. Oren, D., Eldar, Y. C. & Segev, M. Weak measurements compressed sensing quantum state tomography. in *Lasers and Electro-Optics (CLEO), 2016 Conference on* 1–2 (2016).

36. Tian, L., Lee, J., Oh, S. B. & Barbastathis, G. Experimental compressive phase space tomography. *Opt. Express* **20,** 8296 (2012).

37. Rivenson, Y., Stern, A. & Javidi, B. Overview of compressive sensing techniques applied in holography [Invited]. *Appl. Opt.* **52,** A423 (2013).

38. Katz, O., Bromberg, Y. & Silberberg, Y. Compressive ghost imaging. *Appl. Phys. Lett.* **95,** 131110 (2009).

39. Mishali, M. & Eldar, Y. Sub-Nyquist Sampling. *IEEE Signal Process. Mag.* **28,** 98–124 (2011).

40. Gazit, S., Szameit, A., Eldar, Y. C. & Segev, M. Super-resolution and reconstruction of sparse sub-wavelength images. *Opt. Express* **17,** 23920–46 (2009).

41. Shechtman, Y., Eldar, Y. C., Szameit, A. & Segev, M. Sparsity based sub-wavelength imaging with partially incoherent light via quadratic compressed sensing. *Opt. Express* **19,** 14807–22 (2011).

42. Szameit, A. *et al.* Sparsity-based single-shot subwavelength coherent diffractive imaging. *Nat. Mater.* **11,** 455–459 (2012).

43. Narimanov, E. E. The resolution limit for far-field optical imaging. in *CLEO: 2013* QW3A.7 (OSA, 2013). doi:10.1364/CLEO_QELS.2013.QW3A.7

44. Shechtman, Y. *et al.* Sparsity-based super-resolution and phase-retrieval in waveguide arrays. *Opt. Express* **21,** 24015 (2013).

45. Shechtman, Y. *et al.* Phase Retrieval with Application to Optical Imaging: A contemporary overview. *IEEE Signal Process. Mag.* **32,** 87–109 (2015).

46. Sidorenko, P. *et al.* Sparsity-based super-resolved coherent diffraction imaging of one-dimensional objects. *Nat. Commun.* **6,** 8209 (2015).

47. Mutzafi, M. *et al.* Sparsity-based Ankylography for Recovering 3D molecular structures from single-shot 2D scattered light intensity. *Nat. Commun.* **6,** 7950 (2015).

48. Sidorenko, P. & Cohen, O. Single-shot ptychography. *Optica* **3,** 9 (2016).

49. Oren, D., Shechtman, Y., Mutzafi, M., Eldar, Y. C. & Segev, M. Sparsity-based recovery of three-photon quantum states from two-fold correlations. *Optica* **3,** 226–232 (2016).

50. Reck, M., Zeilinger, Bernstein & Bertani. Experimental realization of any discrete unitary operator. *Phys. Rev. Lett.* **73,** 58–61 (1994).

51. Russell, N. J., Chakhmakhchyan, L., O'Brien, J. L. & Laing, A. Direct dialling of Haar random unitary matrices. (2015). doi:10.1088/1367-2630/aa60ed

52. Weiner, A. M. *Ultrafast optics*. (Wiley, 2009).

53. Smithey, D. T., Beck, M., Raymer, M. G. & Faridani, A. Measurement of the Wigner distribution and the density matrix of a light mode using optical homodyne tomography: Application to squeezed states and the vacuum. *Phys. Rev. Lett.* **70,** 1244–1247 (1993).

54. Mandel, L. & Wolf, E. *Optical coherence and quantum optics*. (Cambridge university press, 1995).

55. Sperling, J., Vogel, W. & Agarwal, G. S. True photocounting statistics of multiple on-off detectors. *Phys. Rev. A* **85,** 23820 (2012).

56. Heilmann, R. *et al.* Harnessing click detectors for the genuine characterization of light states. *Sci. Rep.* **6,** 19489 (2016).

57. Titchener, J. G., Solntsev, A. S. & Sukhorukov, A. A. Two-photon tomography using on-chip quantum walks. *Opt. Lett.* **41,** 4079 (2016).

58. Keshavan, R. H., Montanari, A. & Oh, S. Matrix completion from noisy entries. *J. Mach. Learn.*

*Res.* **11,** 2057–2078 (2010).

59. Keshavan, R. H., Montanari, A. & Oh, S. Matrix Completion From a Few Entries. *IEEE Trans. Inf. Theory* **56,** 2980–2998 (2010).

60. Candes, E. & Recht, B. Exact matrix completion via convex optimization. *Commun. ACM* **55,** 111–119 (2012).

61. Fazel, M., Hindi, H. & Boyd, S. P. Log-det heuristic for matrix rank minimization with applications to Hankel and Euclidean distance matrices. in *Proceedings of the 2003 American Control Conference, 2003.* **3,** 2156–2162 (IEEE).

62. Bentivegna, M. *et al.* Experimental scattershot boson sampling. *Sci. Adv.* **1,** (2015).

63. Flammia, S. T., Gross, D., Liu, Y.-K. & Eisert, J. Quantum tomography via compressed sensing: error bounds, sample complexity and efficient estimators. *New J. Phys.* **14,** 95022 (2012).

64. Roger, T. *et al.* Coherent absorption of N00N states. *arXiv* 1–5 (2016). doi:10.1103/PhysRevLett.117.023601

65. Sperling, J. *et al.* Quantum Correlations from the Conditional Statistics of Incomplete Data. *Phys. Rev. Lett.* **117,** 83601 (2016).

66. Nielsen, M. A. & Chuang, I. Quantum computation and quantum information. (2002).

67. Neumark, M. Spectral functions of a symmetric operator. *Izv. Ross. Akad. Nauk. Seriya Mat.* **4,** 277–318 (1940).

68. Peres, A. *Quantum theory: concepts and methods*. **57,** (Springer Science & Business Media, 2006).

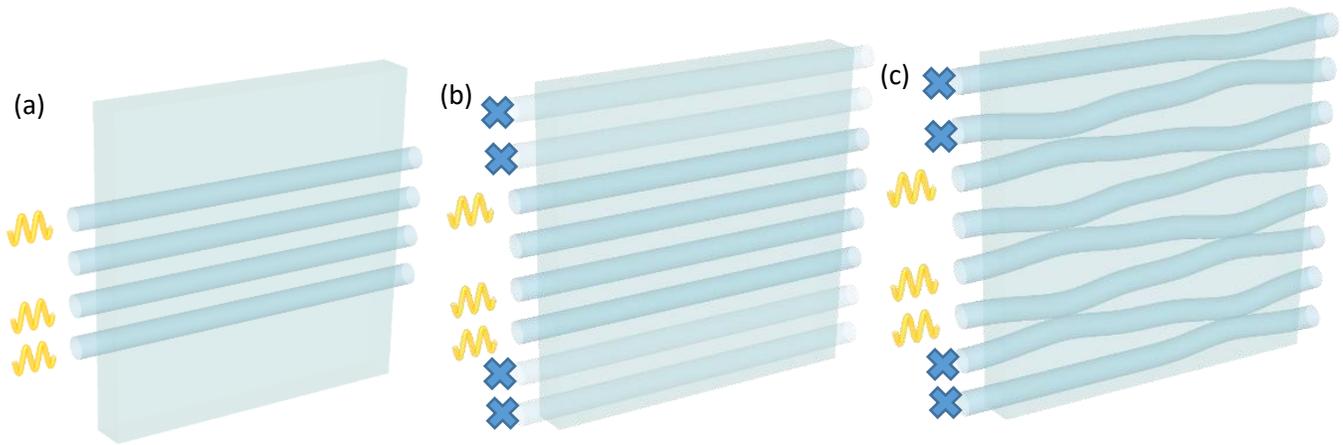

**Figure 1:** (a) A system of $N$ photons in $m$ ports, with $N = 3$ and $m = 4$. The input state is assumed to be sparse, that is close to a pure state. (b) In our scheme, the dimension is increased by addition of an ancilla, taking the form of vacuum ports in a photonic system. (c) The mixing between the degrees of freedom is realized by a random, linear coupler in the larger system.

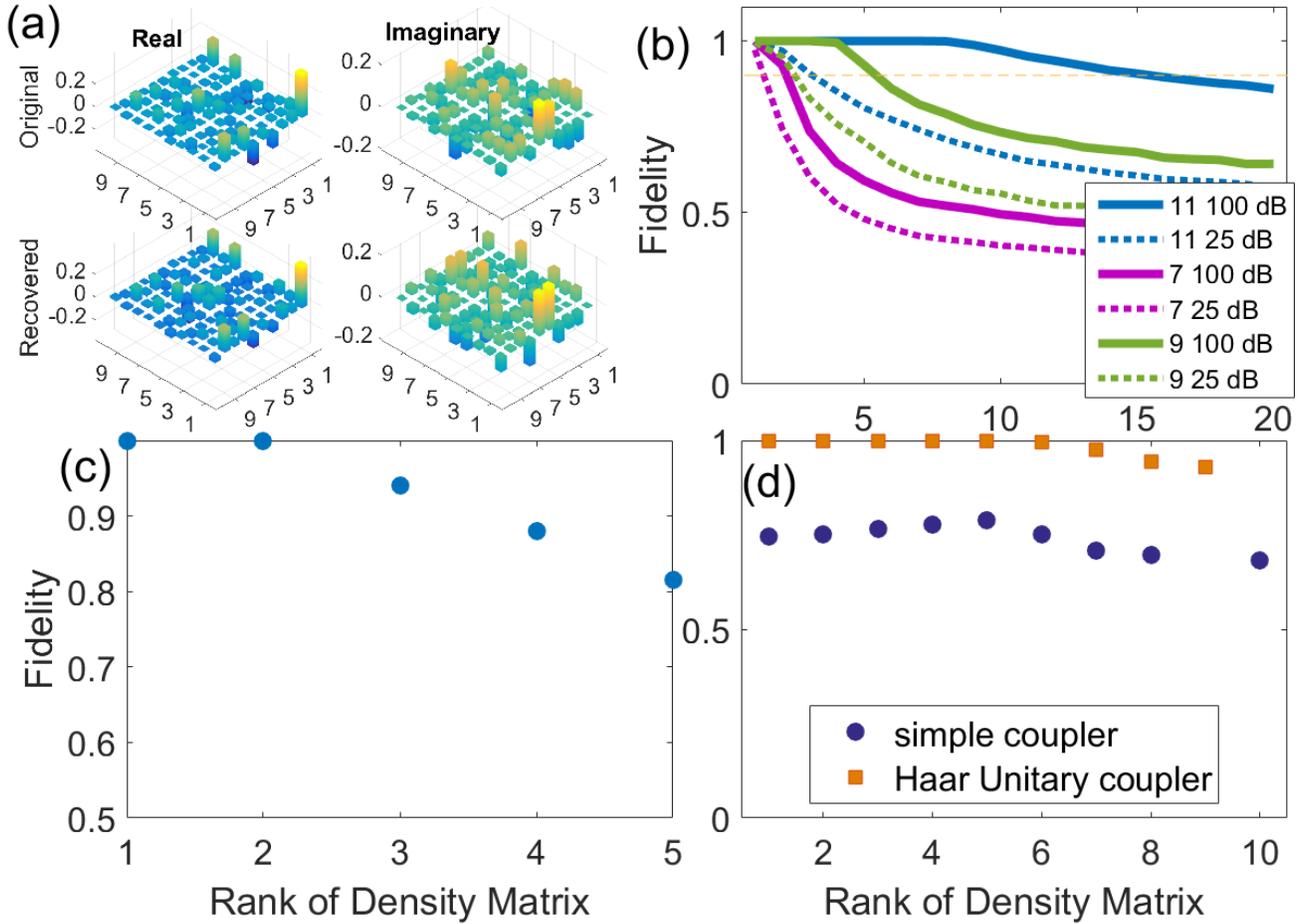

**Figure 2:** (a) A recovery example of a rank-2 density matrix describing $N = 3$ photons in $m = 3$ input ports and $M = 7$ output ports. The state is recovered with fidelity of 0.96. (b) Mean fidelity of the state recovered from measurements of a single observable versus rank of input state. Solid curves: recovery without noise using 7,9,and 11 output ports. Dotted curves: same as the solid curves but with depolarization noise added to the state, and measurement noise of $25 dB$ added to the measurements. The plots show the average over 200 realizations of the density matrix and 10 realizations of the random coupler for each point. The measurements used here are only a portion (21-71%) of the measurements required for full QST. (c) Mean fidelity versus rank of input state, describing $N = 3$ photons in $m = 7$ input ports and $M = 16$ output ports, averaged over 15 realizations of the density matrix. Here, the dimension of the system is large, $d = 84$. The number of measurements in a single setup in this scenario is 11% of the measurements required for full QST. (d) Comparison between a fully mixing coupler (randomly sampled from Haar measure) and a simpler coupler, consisting of identical evanescently coupled waveguides. The simpler coupler fails to allow recovery from a single observable, whereas the Haar coupler performs well up to Rank 6.

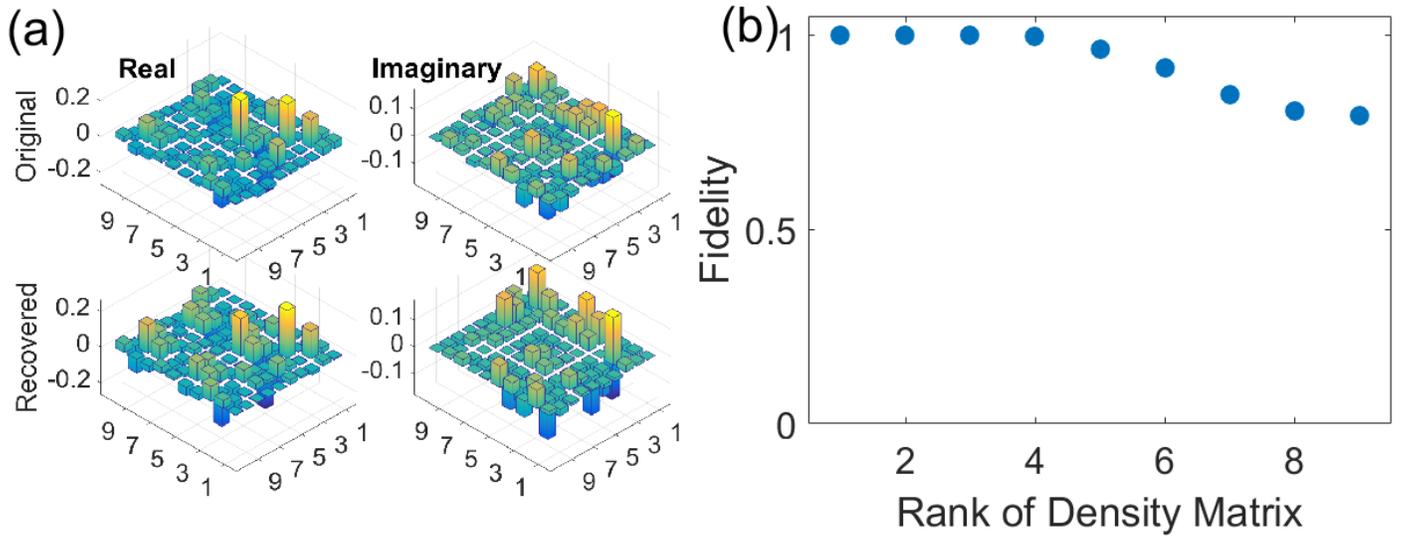

**Figure 3:** (a) A recovery example of a rank-2 density matrix describing $N = 3$ photons in $m = 3$ input ports, $M = 8$ output ports, with $25 dB$ of measurements noise. The state is recovered with $0.93$ fidelity, from measurements using click detectors. (b) Recovery fidelity with click detectors of density matrices of $N = 3$ photons in $m = 4$ output ports and $M = 11$ output ports, corresponding to $41\%$ of the measurements required for QST, in a single setup.

# Quantum State Tomography with a Single Observable –

# Supplementary Material

Dikla Oren[1], Maor Mutzafi[1], Yonina C. Eldar[2], Mordechai Segev[1]

[1]Physics Department and Solid State Institute, Technion, 32000 Haifa, Israel

[2] Electrical Engineering Department, Technion, 32000 Haifa, Israel

## Coupler Description

The linear, passive coupler $U$ of $M$ modes is represented by a Unitary matrix of dimension $M$, $U \in \mathrm{U}(M)$, where $\mathrm{U}(M)$ is the Unitary group of order $M$. The coupler $U$ is sampled from the Haar measure on $\mathrm{U}(M)$, which is the uniform measure on the Unitary group.

The photons entering the coupler evolve according to

$$a_i^\dagger \mapsto \sum_j U_{ij} a_j^\dagger. \qquad \text{(S 1)}$$

Here, $a_i^\dagger$ is the photon creation operator at port $i$. The photon evolution induces a Unitary evolution in the entire space $\mathcal{U} \in \mathrm{U}(D)$, where $D$ is the dimension of the truncated Fock space $D = \binom{M-1+N}{N}$ for $N$ photons. Thus, the density matrix at the input $\rho$ evolves at the output to

$$\rho \mapsto \rho_{\text{out}} = \mathcal{U} \rho \mathcal{U}^\dagger. \qquad \text{(S 2)}$$

## Observable Details

For $M$ ports, $N$ photons, $M > N$, define the following observable

$$A = \sum_{q_1,\ldots,q_N} |1_{q_1}\cdots 1_{q_N}\rangle\langle 1_{q_1}\cdots 1_{q_N}| \quad \text{(S 3)}$$

$$+ 2 \sum_{q_1,\ldots,q_{N-1}} |1_{q_1}\cdots 1_{q_{N-2}} 2_{q_{N-1}}\rangle\langle 1_{q_1}\cdots 1_{q_{N-2}} 2_{q_{N-1}}|$$

$$+ 3 \sum_{q_1,\ldots,q_{N-2}} |1_{q_1}\cdots 1_{q_{N-3}} 3_{q_{N-2}}\rangle\langle 1_{q_1}\cdots 1_{q_{N-3}} 3_{q_{N-2}}|$$

$$+ 4 \sum_{q_1,\ldots,q_{N-2}} |1_{q_1}\cdots 1_{q_{N-4}} 2_{q_{N-3}} 2_{q_{N-2}}\rangle\langle 1_{q_1}\cdots 1_{q_{N-4}} 2_{q_{N-3}} 2_{q_{N-2}}|$$

$$+ 5 \sum_{q_1,\ldots,q_{N-3}} |1_{q_1}\cdots 1_{q_{N-4}} 4_{q_{N-3}}\rangle\langle 1_{q_1}\cdots 1_{q_{N-4}} 4_{q_{N-3}}| + \cdots$$

$$= \sum_{\{n\}^i} i |\{n\}^i\rangle\langle\{n\}^i|.$$

Here, $\{n\}^i = (n_1^i \cdots n_M^i), \sum_q n_q^i = N$ is the $i$th configuration of $N$ photons in $M$ ports. The different projections are obtained from all Fock states of $M$ ports, $N$ photons. It is indeed an observable, since $A^\dagger = A$. Experimentally, each projection corresponds to a different $N$-fold correlation measurement. For example, $|1_{q_1}\cdots 1_{q_N}\rangle\langle 1_{q_1}\cdots 1_{q_N}|$ corresponds to measuring a single photon in port $q_1$, a single photon in port $q_2$ and so on.

The addition of the linear coupler $U \in U(M)$ is represented by a Unitary matrix of dimensions $M \times M$ (see Coupler section in this supplementary), dictating an evolution of the photons as in eq. (S 1).

To find the probability of each measurement outcome, we can use the evolution of the density matrix, as in eq. (S 2). Alternatively, we can fix the input state and evolve the observable

$$A' = \mathcal{U}^\dagger A \mathcal{U}. \quad \text{(S 4)}$$

The probability for each $N$-fold correlation is

$$p(|\{n\}^i\rangle|\rho) = \text{Tr}(\rho \mathcal{U}^\dagger |\{n\}^i\rangle\langle\{n\}^i|\mathcal{U}) = \langle\{n\}^i|\mathcal{U}\rho\mathcal{U}^\dagger|\{n\}^i\rangle, \quad i \in \{1,\ldots,D\}. \quad \text{(S 5)}$$

Since these are all correlations after propagation in a single coupler $U$, they require a single experimental setup.

## Recovery Methods

The problem of recovering the density matrix from measurements can be described as finding the density matrix $\rho$ such that the measurement outcomes derived from $\rho$ fit the experimental results. If we denote the linear transformation of the density matrix due to the observables being measured by $\mathcal{A}_i\colon \mathbb{C}^{d\times d} \to \mathbb{R}_+^d$, $\mathcal{A}_i(\rho) = y_i, i \in \{1, \ldots, d\}$, where $d$ is dimension of the system, then our problem can be described mathematically as: Find $\rho \in \mathbb{C}^{d\times d}$ such that $\rho^\dagger = \rho, \rho \geq 0, \mathrm{Tr}\rho = 1$ and $\mathcal{A}_i(\rho) = y_i$ for $i = 1, \ldots, d$. The first constraints result from $\rho$ being a density matrix, while the rest arise from the requirement to conform to the measurements. This problem can be formulated as the optimization problem

$$\min_\rho \sum_{i=1}^{d} \|\mathcal{A}_i(\rho) - y_i\|_2^2 \qquad (S\ 6)$$

$$\text{subject to } \rho^\dagger = \rho, \rho \geq 0, \mathrm{Tr}\rho = 1.$$

In our problem, we wish to find the density matrix at the input $\rho_0$, such that the derived measurements after the coupler, with the addition of the ancilla, conform to the measured data. Since we use a single observable in the larger system of dimension $D$, the measurements are described by the linear transformation $\mathcal{A}\colon \mathbb{C}^{D\times D} \to \mathbb{R}_+^D$, $\mathcal{A}(\rho_0 \otimes \rho_{\text{ancilla}}) = y$ with the $y_j$ given by eq. (S 5).

In our scheme, we use prior knowledge, namely that the state is close to a pure state or can be approximated by one. In terms of the density matrix, the prior knowledge takes the form of the density matrix having a small number of nonzero eigenvalues. In other words, the density matrix has low rank, or can be approximated by a low rank matrix. To harness this knowledge for the purpose of state recovery, we consider the following optimization problem

$$\min_{\rho_0} \mathrm{rank}(\rho_0) \qquad (S\ 7)$$

$$\text{subject to } \rho_0^\dagger = \rho_0, \rho_0 \geq 0, \mathrm{Tr}\rho_0 = 1,$$

$$\mathcal{A}(\rho_0 \otimes \rho_{\text{ancilla}}) = y.$$

Here, $\rho_0$ is the state at the input that we wish to recover, $\rho_{\text{ancilla}}$ is the state of the ancilla at the input (known to us) and $y$ are the experimentally measured data. With the realistic addition of experimental noise on the measurements, we can modify (S 7) to

$$\min_{\rho_0} \mathrm{rank}(\rho_0) \qquad (S\ 8)$$

$$\text{subject to } \rho_0^\dagger = \rho_0, \rho_0 \geq 0, \mathrm{Tr}\rho_0 = 1,$$

$$\|\mathcal{A}(\rho_0 \otimes \rho_{\text{ancilla}}) - y\|_2^2 \leq \epsilon,$$

where $\epsilon$ is related to the amount of measurement noise. The problem formulated in eq. (S 8) is not convex due to the rank objective. To approximate the solution, we use the LogDet heuristic [1], in which the non-convex rank function is replaced by a surrogate function, the log of the determinant, which promotes low rank solutions. The LogDet function is then linearized at the vicinity of a proposed solution to yield an iterative algorithm. In the $k$th iteration, the optimization problem to be solved is

$$\min_{X_k} \text{Tr}(X_{k-1} + \delta I)^{-1} X_k \qquad (S\ 9)$$
$$\text{subject to } X_k \geq 0, \text{Tr}(X_k) = 1,$$
$$\|\mathcal{A}(\rho_0 \otimes \rho_{\text{ancilla}}) - y\|_2^2 \leq \epsilon,$$

Here, $X_k$ is the matrix we are looking for in iteration $k$, $X_{k-1}$ is the matrix found in the previous iteration, and $\delta$ is a small regularization parameter. Once $X_k, X_{k-1}$ are close to each other in the Frobenius norm sense, no further iterations are preformed, and the low rank density matrix is $X_k$.

A second approach to solving (S 8) is to drop the low-rank requirement and solve the problem

$$\min_{\rho_0} \|\mathcal{A}(\rho_0 \otimes \rho_{\text{ancilla}}) - y\|_2^2 \qquad (S\ 10)$$
$$\text{subject to } \rho_0^\dagger = \rho_0, \rho_0 \geq 0, \text{Tr}\rho_0 = 1.$$

In this case, the prior knowledge is not used explicitly. However, for certain measurement matrices, the mere structure of the density matrix, namely the positive-definiteness, results in a unique low rank solution [2,3]. Thus, in these cases, the low rank solution is found regardless of the algorithm used. A comparison between LogDet (eq. (S 9)) and the constrained least squares problem (eq. (S 10)) is presented in Fig. 1S. The results are very similar, with the LogDet somewhat superior.

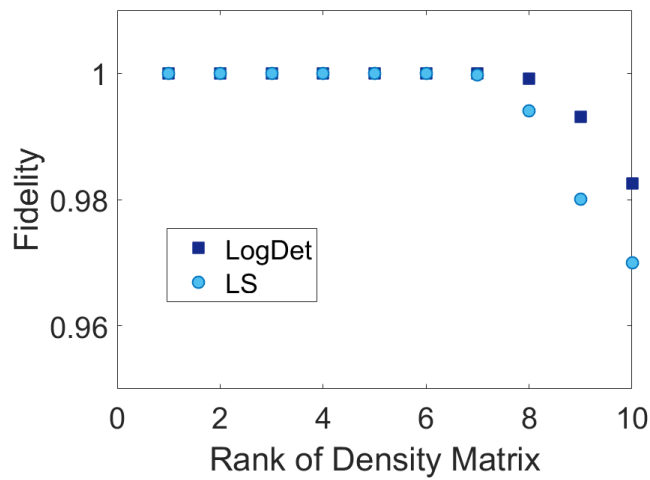

**Figure 1S:** A comparison between the recovery fidelity for $N = 3$ photons in $m = 4$ input ports, $M = 11$ output ports, using LogDet (eq. (S 9)) and constrained LeastSquares (eq. (S 10)).

## Sampling Mixed States

The problem of sampling quantum states from the entire Hilbert space has theoretical importance, as well as direct consequences on benchmarking of quantum circuits and quantum state recovery methods. For pure states, described by density matrices with rank 1, there exists a unique uniform measure, the Haar measure, on the Unitary group of dimensions corresponding to the dimension of the system. This measure is invariant to Unitary transformations, thus all pure states are equivalent.

While the question of sampling pure states has a clear answer, the answer to the corresponding problem with mixed states remains elusive [4]. The density matrix, as a positive semidefinite matrix with trace 1, can be described by its spectral decomposition $\rho = \sum_i \lambda_i |i\rangle\langle i|$, where $\lambda_i$ are the eigenvalues and $|i\rangle$ are the eigenvectors. Thus, a measure $\mu$ on the set of density matrices depends on the eigenvalues and eigenvectors $\mu = \mu(\lambda_1, \ldots, \lambda_D; |1\rangle, \ldots, |D\rangle)$.

In the spirit of the Unitary invariance satisfied by the uniform measure on pure states, we can define measures on the set of density matrices which are product measures on the eigenvalues and eigenvectors [5]

$$\mu = \mu_\lambda(\lambda_1, \ldots, \lambda_D) \times \mu_i(|1\rangle, \ldots, |D\rangle), \qquad (S\ 11)$$

where the measure on the eigenvectors $\mu_i$ is the Haar measure. Other measures include, for example, the measure induced by pure states in a larger system [4]. They are sampled uniformly, and then a part of the system is traced over to produce a mixed state of the original dimension.

As explained in the main text, to demonstrate our quantum state recovery method, we preform recovery simulations. In these simulations, a large number (typically several hundreds) of density matrices are generated, they evolve in the system of a single observable, and the measurement outcomes are calculated according to eq. (S 5). We then add noise to the measurements, and apply our method to recover the state at the input of the system. The recovered density matrix is then compared to the original one. Thus, we need to sample a large number of density matrices describing mixed states.

Due to the appeal and physical motivation of the Unitary invariance, we choose to sample from a product measure on the eigenvalues and eigenvectors (eq. (S 11)), where we take the eigenvectors measure to be

Haar, leading to invariance of the measure under Unitary evolution. The measure on the eigenvalues is taken to be the uniform measure on the $r - 1$ dimensional unit simplex, where $r$ is the desired rank of the density matrix.

## Dependence on the Realization of the Coupler

The mixing component in our scheme is realized in the optical setting by a linear random coupler. The coupler is sampled from the Haar measure on the corresponding Unitary group. It is natural to ask how strongly, if at all, the specific realization of the coupler affects the ability to recover the quantum state. To answer this question, we sample 30 different couplers from the same distribution and use them to recover a single density matrix of rank 2, describing 3 photons in 4 input modes, with 10 output modes (thus ~0.55 of the measurements required for tomography are used). The results are shown in Fig. 2S(a). The mean fidelity in this scenario is 0.996. As evident from the figure, the fidelity does not depend on the specific realization of the coupler. In Fig. 2S(b), the fidelity for each coupler is averaged over 30 density matrices, again showing no dependence on the specific realization of the coupler, as long as it is sampled uniformly.

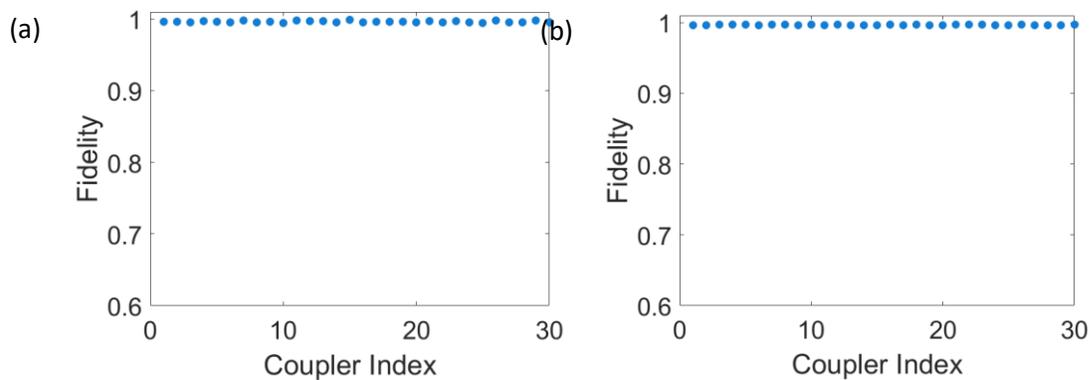

**Figure 2S:** (a) Fidelity obtained with different couplers, a single density matrix. (b) Fidelity obtained with different coupler, averaged over 30 density matrices.

## A POVM Set of the Measurements

The ancilla added to the original system increases the dimension of the system, resulting in addition of measurements. However, the evolution in the system has to couple between the ancilla and the original system to indeed yield more measurements. This can be shown using a Positive Operator Valued Measurements (POVM) formulation of the measurements performed in our scheme. POVM measurements are generalized quantum measurements, described by a set of operators $\{E_i\}_{i=1}^n$ satisfying $E_i \geq 0, \sum_i E_i = I$. The probability of each outcome is $p(i|\rho) = \text{Tr}(E_i \rho)$. They are called generalized measurements since the operators $E_i$ do not have to be orthogonal to each other, unlike the case of ordinary measurements in quantum mechanics, called projective measurements.

The physical realization of a POVM set $\{E_i\}_{i=1}^n$ is obtained by extending the original Hilbert space using an ancilla, and performing projective measurements in the larger space, as described in Neumark Theorem [6,7]. In the process of realizing the POVM set, we start with a set of operators on the original Hilbert space as described earlier, and we obtain an ancilla and a set of projective measurements in the combined Hilbert space of the original system and the ancilla. In our scheme, we describe a set of projective measurements, namely the spectral decomposition of (S 4), and an ancilla $\rho_{\text{ancilla}} = |0\rangle\langle 0|$. By deriving the corresponding POVM set on the Hilbert space of our original density matrix $\rho_0$, we can investigate the interplay between the coupler and the ancilla.

As in the main text, we denote the dimension of the original system as $d$, that of the extended system (with the ancilla) as $D$, such that $\rho_0 \in \mathbb{C}^{d \times d}$, $\rho_{\text{in}} = \rho_0 \otimes \rho_{\text{ancilla}} \in \mathbb{C}^{D \times D}$. The original system consists of ports $i \in I_{\text{original}}$, $|I_{\text{original}}| = m$, while the ancilla consists of ports $i \in I_{\text{ancilla}}$, and $|I_{\text{original}} \cup I_{\text{ancilla}}| = M$.

The projective measurements are $P_i = w_i w_i^\dagger, i \in \{1, \dots, D\}$, where $w_i = \mathcal{U}|\{n\}^i\rangle$. There are indeed $D$ such orthogonal vectors, since the Fock states $\{|\{n\}^i\rangle\}_{i=1}^D$ form an orthonormal basis, and the evolution operator $\mathcal{U}$ is unitary.

The POVM elements $E_i$ are related to the $w_i$ by a projection on the original subspace [7]

$$E_i = \text{Tr}_{\text{ancilla}}(v_i v_i^\dagger) \quad i \in \{1, \dots, D\} \quad \text{(S 12)}$$

$$v_i = \text{P}_{\text{original}}(w_i)$$

where the projection onto the original space is defined in our case as

$$\text{P}_{\text{original}}\left(\sum_{i=1}^D \beta_i |\{n\}^i\rangle\right) = \sum_{i \in I} \beta_i |\{n\}^i\rangle \quad \text{(S 13)}$$

$$I = \{i \in \{1, \dots, D\} \mid n_q^i = 0 \ \forall q \in I_{\text{ancilla}}\}.$$

That is, the projection eliminates all the coefficients of the vectors in the large space having a nonzero number of photons in the ancilla ports. For the vectors defining the projections, we have

$$v_i = \mathrm{P}_{\mathrm{original}}(w_i) = \sum_{j \in I} \mathcal{U}_{ij} |\{n\}^j\rangle \tag{S 14}$$

and therefore, the POVM operators are

$$E_i = \mathrm{Tr}_{\mathrm{ancilla}} \left( \sum_{j,j' \in I} \mathcal{U}_{ij} \mathcal{U}^\dagger_{j'i} |\{n\}^j\rangle\langle\{n\}^{j'}| \right) \tag{S 15}$$

These are indeed operators on the desired space, since $\forall j \in I$, we have zero photons in the ancilla ports. Therefore, we can write

$$E_i = \mathrm{Tr}_{\mathrm{ancilla}} \left( \sum_{j,j' \in I} \mathcal{U}_{ij} \mathcal{U}^\dagger_{j'i} |n_1^j \cdots n_m^j\rangle\langle n_1^{j'} \cdots n_m^{j'}| \otimes |0\rangle\langle 0| \right) \tag{S 16}$$

$$= \sum_{j,j' \in I} \mathcal{U}_{ij} \mathcal{U}^\dagger_{j'i} |n_1^j \cdots n_m^j\rangle\langle n_1^{j'} \cdots n_m^{j'}|.$$

Here, $i \in \{1, \ldots, D\}$ and $j, j' \in I$. The operators $E_i$ form a resolution of the identity

$$\sum_{i \in \{1,\ldots,D\}} E_i = \sum_{i \in \{1,\ldots,D\}} \sum_{j,j' \in I} \mathcal{U}_{ij} \mathcal{U}^\dagger_{j'i} |n_1^j \cdots n_m^j\rangle\langle n_1^{j'} \cdots n_m^{j'}| \tag{S 17}$$

$$= \sum_{j,j' \in I} \sum_{i \in \{1,\ldots,D\}} \mathcal{U}_{ij} \mathcal{U}^\dagger_{j'i} |n_1^j \cdots n_m^j\rangle\langle n_1^{j'} \cdots n_m^{j'}|$$

$$= \sum_{j,j' \in I} \delta_{jj'} |n_1^j \cdots n_m^j\rangle\langle n_1^{j'} \cdots n_m^{j'}| = \sum_{j \in I} |n_1^j \cdots n_m^j\rangle\langle n_1^j \cdots n_m^j| = I.$$

They are also rank-1 operators, since

$$E_i = |q_i\rangle\langle q_i| \tag{S 18}$$

$$|q_i\rangle = \sum_{j \in I} \mathcal{U}_{ij} |n_1^j \cdots n_m^j\rangle,$$

therefore they are positive semidefinite $\langle\psi|E_i|\psi\rangle = |\langle\psi|q_i\rangle|^2 \geq 0 \ \forall |\psi\rangle$. Thus, the set $\{E_i\}_{i=1}^D$ forms a POVM.

The measurements-state relation can be formulated in the following manner

$$y = \mathcal{A} \rho_{CS}. \tag{S 19}$$

Here, $y \in \mathbb{R}_+^D$, $\mathcal{A} \in \mathbb{C}^{D \times d^2}$ and $\rho_{CS} \in \mathbb{C}^{d^2}$ is the original density matrix in vector form, obtained by column stacking. The matrix $\mathcal{A}$ describes the measurements. Since $y_i = \mathrm{Tr}(E_i \rho)$, by using the fact that the $E_i$ are

rank-1, we can obtain the elements of $\mathcal{A}$. Define a set of $D$ matrices $B^k \in \mathbb{C}^{d \times d}, k \in \{1, \ldots, D\}$ with elements

$$B^k_{ij} = \mathcal{U}_{ki} \mathcal{U}^\star_{kj}. \qquad (S\ 20)$$

Here, $\mathcal{U}$ is the evolution operator. Then the $k$th row of $\mathcal{A}$ is the row stacking of $B^k$. To estimate the usability of the measurements, we sampled 100 random linear couplers from the Haar measure of varying dimensions, and plotted in Fig. 3S the average rank of the measurement matrix $\mathcal{A}$ versus the number of measurements $D$. The original system dimension is $d = 20$. As evident in the figure, the measurement matrix has full row rank, up to the point where $D > d^2$.

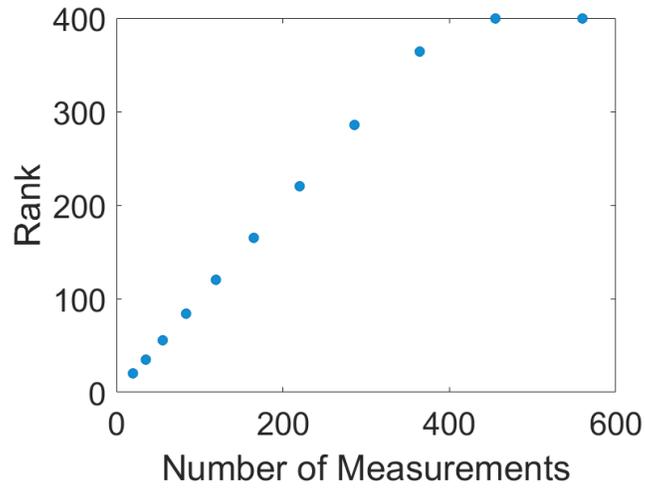

**Figure 3S:** Average rank of $\mathcal{A}$ (eq. (S 19)) versus the number of measurements $D$.

Notice that for an interaction-less coupler $\mathcal{U} = \mathcal{U}_{\text{original}} \otimes \mathcal{U}_{\text{ancilla}}$, we have in eq. (S 14)

$$v_i = \sum_{j \in I} \mathcal{U}_{ij} |n_1^j \cdots n_m^j\rangle \otimes |0\rangle. \qquad (S\ 21)$$

However, due to the tensor product, the $v_i$ are linearly dependent. In that case, there are only at most $d$ linearly independent $v_i$, and hence the number of measurements is lower than $d$.

# References

1. M. Fazel, H. Hindi, and S. P. Boyd, "Log-det heuristic for matrix rank minimization with applications to Hankel and Euclidean distance matrices," in *Proceedings of the 2003 American Control Conference, 2003.* (IEEE, n.d.), Vol. 3, pp. 2156–2162.

2. M. Wang, W. Xu, and A. Tang, "A Unique "Nonnegative" Solution to an Underdetermined System: From Vectors to Matrices," IEEE Trans. Signal Process. **59**, 1007–1016 (2011).

3. A. Kalev, R. L. Kosut, I. H. Deutsch, J. Řeháček, and Z. Hradil, "Quantum tomography protocols with positivity are compressed sensing protocols," npj Quantum Inf. **1**, 15018 (2015).

4. K. Zyczkowski and H.-J. Sommers, "Induced measures in the space of mixed quantum states," J. Phys. A. Math. Gen. **34**, 7111 (2001).

5. K. Życzkowski, "Volume of the set of separable states. II," Phys. Rev. A **60**, 3496–3507 (1999).

6. M. Neumark, "Spectral functions of a symmetric operator," Izv. Ross. Akad. Nauk. Seriya Mat. **4**, 277–318 (1940).

7. A. Peres, *Quantum Theory: Concepts and Methods* (Springer Science & Business Media, 2006), Vol. 57.